\documentclass[aps,prl,twocolumn,showpacs,superscriptaddress,groupedaddress]{revtex4-1}  
\usepackage{epsfig}
\usepackage{graphicx}
\usepackage{natbib}

\begin{document}

\author{Somayeh Farhadi} \author{Robert P Behringer} \author{Alex Z Zhu}
\affiliation{Department of Physics and Center for Nonlinear and Complex Systems, Box 90305, Duke University, Durham, NC 27708}
\date{\today} \title{Stress relaxation for granular materials near
  Jamming under cyclic compression}

\renewcommand{\textfraction}{0.05}
\renewcommand{\topfraction}{0.95}
\renewcommand{\bottomfraction}{0.95}
\setcounter{bottomnumber}{4} 
\setcounter{topnumber}{4}
\renewcommand{\floatpagefraction}{0.95}

\newcommand{\ea}{{\it et al.}}

\begin{abstract}        
We have explored isotropically jammed states of semi-2D granular
materials through cyclic compression. In each compression cycle,
systems of either identical ellipses or bi-disperse disks, transition
between jammed and unjammed states. We determine the evolution of the
average pressure, $P$, and structure through consecutive jammed
states. We observe a transition point, $\phi_m$, above which $P$
persists over many cycles; below $\phi_m$, $P$ relaxes slowly. The
relaxation time scale associated with $P$ increases with packing
fraction, while the relaxation time scale for collective particle
motion remains constant. The collective motion of the ellipses is
hindered compared to disks, due to the rotational constraints on
elliptical particles.
\end{abstract}

\maketitle {\it Introduction:} Particle systems near jamming exhibit
several signature features, including dynamical slowing down, and
heterogeneous dynamics\cite{Dynamical_Heterogeneities,lacevic04}.
Systems of interest include colloids, molecular glass formers, and
granular materials\cite{keys07,Dauchot05,glotzer00,coulais12}.
Although all these systems can jam, granular materials, which we
consider here, have the experimental advantage of accessibility at the
particle scale.  We use this feature to explore the dynamics of
isotropically driven disordered materials near jamming.

Several aspects distinguish studies of these disparate systems,
including excitation mechanisms, and interparticle interactions.  In
molecular and colloidal systems, temperature provides homogeneous
driving. In granular systems, temperature is an irrelevant variable,
and driving must be provided externally.  In past granular studies,
driving came from vibration or
tapping\cite{richard05,Philippe02,Lechenault08}, or by
shear\cite{howell99}.  Vibration and tapping usually involve energy
input on rapid time scales and in ways that may not be isotropic and
uniform.  Shear strain can be applied on any time scale, but it is
anisotropic, and not necessarily homogeneous; e.g., shear failures are
often localized\cite{Jaeger96,Jaeger92}.  The first point of the
present experiments is to understand the effect of jamming on granular
systems when the driving mechanism is (relatively) uniform, isotropic
and on slow time scales.

Previous studies of spatio-temporal granular dynamics near jamming
have typically involved spherically symmetric particles: disks in 2D,
spheres in 3D.  In recent experiments, we showed that tangential
forces for frictional particles significantly changes stable states
near jamming\cite{majmudar05}, and helps stabilize the granular
network (`force chains').  If friction stabilize granular networks by
limiting rotation, it is natural to probe the possibly similar role
played by geometry, and this is the second question that we address
here.  As we show below, although both types of systems slow down
under cyclic driving as the density grows, the characteristic time
scales for ellipses is significantly greater than for disks.

In the present experiments, we cyclically and isotropically
compress/expand our granular systems by small amounts, starting from
a packing fraction $\phi$, just below isotropic jamming, and
compressing to a $\phi$ that is above the isotropic jamming. We
measure the mean pressure, $P$, and the collective dynamics in the
most compressed states for very large numbers of cycles. Under this
protocol, the system may slowly find more compact configurations.  The
time associated with this evolution becomes large as $\phi$ for the
compressed state grows, whereas the time scale for the evolution of
inhomogeneities of particle motion remains roughly constant.  We
emphasize that the most compressed $\phi$ during a cycle is always
above isotropic jamming, e.g. $\phi_c \simeq 0.84$ for disks and
$\phi_c \simeq 0.91$ for ellipses\cite{farhadi-thesis12}.

\begin{figure}
\centering
{\includegraphics[scale=0.38]{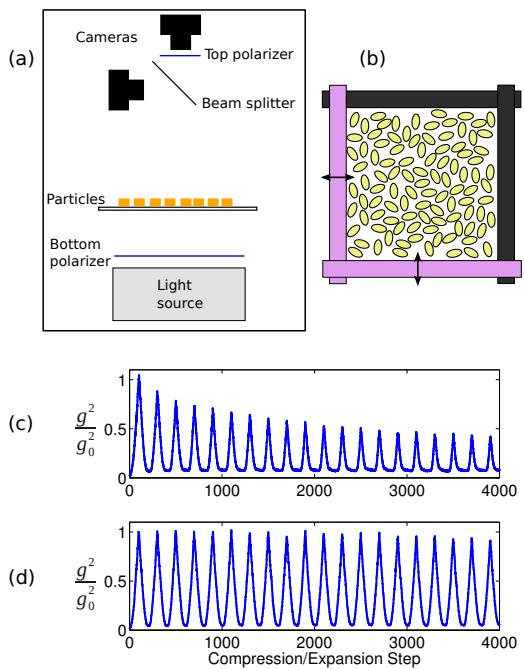}}
\caption{Schematics of the biax. a) Particles are confined inside a
  square region. The grey walls are stationary and the colored walls,
  which are attached to independent linear motors, move. b) Side view
  of the setup. c,d) Time series of global $g^{2}$ for two sets of
  cyclic compression of ellipses performed in different packing
  fraction intervals of c) [0.868, 0.896], and d) [0.906, 0.937]. The
  values of $g^2$ are normalized by $g^2$ at the most compressed state
  of the first cycle. The system is strained (e.g. compressed) by
  $\%1.6$ by a series of much smaller quasi-static step in each
  compression.}
\label{fig:Biaxial_schematics}
\end{figure}

{\it Experiment:} 
The experiments consisted of cyclic isotropic compression of quasi-2D systems of bidisperse disks and systems of identical ellipses with aspect ratio $1.85$. A schematic of the setup is
shown in Fig.~\ref{fig:Biaxial_schematics}. The particles were
confined in a square container, a biax, where two of the confining
walls were stationary, while the other two were displaced using linear
motors. The distance between opposing pairs of walls has a spatial
resolution of $\sim 10^{-6}$m.  The number of particles was kept
constant at 2400 in all the experiments, regardless of particle
type, and the packing fraction was varied by changing the area of the
confining square of the biax \cite{majmudar05}.  The packing
fractions, $\phi$, of the fully compressed states were chosen above
the isotopic jamming point (point J).  The particles were
photoelastic, which allowed us to measure the local pressure acting on
each particle. Before starting any compression cycle, the system was
prepared in a stress-free state.  It was then quasi-statically
compressed via many small strain steps (about $0.016\%$) for a typical
total volumetric strain of $3.2\%$.  Next, it was quasi-statically
expanded to its initial strain-free state. This process was repeated
for multiple cycles. Some initial runs were made for which the system
was imaged at every strain step for multiple compressions and
relaxations. For longer numbers of compression cycles (up to $~\sim1000$ cycles), the system was strained as above, but imaged only once
at the maximum compressive strain for each cycle. Before
imaging at
this strain extremum, the system was allowed to
relax. The imaging was carried out using two synchronized digital
cameras, one of which recorded a polarized image which yielded the
photoelastic response of the system, while the other recorded a normal
(unpolarized) image.  These two images enabled us to measure the local
stress on each particle, and to track the particles, including their
centers, and orientations (for ellipses).

The quasi-2D particles, either circular (disks) or ellipses, are
machined from sheets of Vishay polymer PSM-4 that are $0.635$~cm
thick.  The semi-minor axis of the ellipses is $b \simeq 0.25$ cm, and
the aspect ratio is $1.85$. The system of disks is bi-disperse, where the
ratio of small-to-large particles is kept at about $4.5:1$. The radius
of small disks is $r_s \simeq 0.38$ cm, and the radius of larger
particles is $r_l \simeq 0.44$ cm.  The particles rest on a horizontal
Plexiglas sheet (Fig.~\ref{fig:Biaxial_schematics}a) that has been
lubricated by a layer of fine powder. As shown in
Fig.~\ref{fig:Biaxial_schematics}a, a circularly polarized beam passes
through the Plexiglas sheet and the particles from below, and then
through a beam splitter placed over the setup. The beam splitter
provides two identical images, one of which is viewed, without a
polarizer by the horizontal camera, while the other, passes through a
crossed circular polarizer (with respect to the bottom polarizer) and
is viewed by the top camera. The two cameras acquire images
simultaneously. As a result, one camera records the photoelastic
response of the system (polarized image), and the other camera yields
a direct image (normal image) of the particles for tracking. We
extracted the local stress acting on each particle, which is encoded
in the photoelastic response, using an established empirical measure
which we call $g^2$.  This quantity is the gradient squared of the
transmitted photoelastic image intensity integrated over the pixels
associated with each particle\cite{howell99}. Before computing $g^2$,
we first filter the photoelastic images, leaving only the green
channel of the original color image, which corresponds to the optimum
color response of the polarizers.

\begin{figure}
\centering
{\includegraphics[scale=0.4]{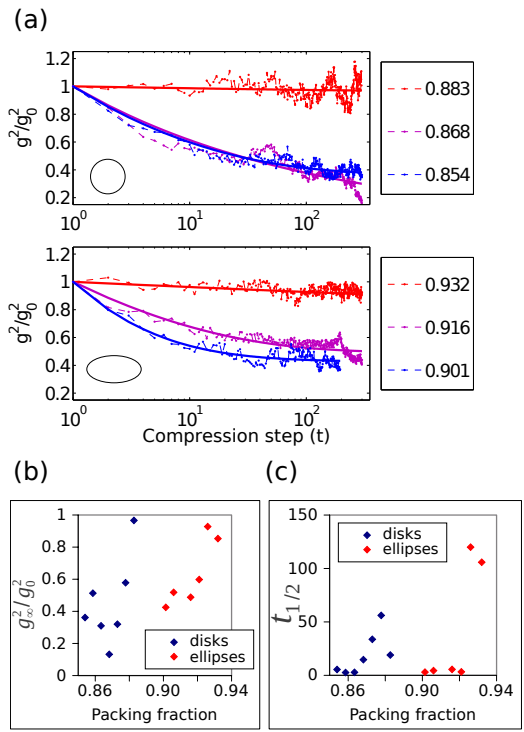}}
\caption{a) Time series of global $g^{2}$ for representative systems
  of disks, and ellipses under maximum compression. Colors represent
  packing fractions, $\phi$, at the most compressed state of each
  cycle. Each step is prepared by compressing the system with a strain
  value of $\%1.6$. Here, $g_{0}^{2}$ is the value of $g^{2}$ at the
  initial compressed state. b) Data for $g_{\infty}$, determined
  from fit data using the parameters $a$, and $t_0$ to time series
  data such as those in Fig.\ref{fig:gsqr_time_series_2}a to the
  equation: $\frac{g^2}{g^2_0}=(\frac{t+t_0}{t(t_0+1)})^{a}$. c) Data
for the relaxation time for $\frac{g^2}{g^2_0}$ to relax halfway
between its initial value of 1, and $g_{\infty}$.}
\label{fig:gsqr_time_series_2}
\end{figure}

{\it Global stress response:}
Fig.~\ref{fig:Biaxial_schematics}(c,d) show time series of the global
stress of the ellipses for two packing fraction intervals
(i.e. minimum and maximum $\phi$'s), which are obtained by
incrementally compressing/expanding the system over a number of
cycles. The global stress is found by averaging $g^2$ over all
particles in each image at a given strain. In the plots of
Fig.~\ref{fig:Biaxial_schematics}(c,d), the values of $g^{2}$ are
normalized by the integrated $g^{2}$ at the most compressed state of
the first cycle ($g^{2}_{0}$), which always corresponds to a density
that is above isotropic jamming point. For the lower packing fraction,
Fig.~\ref{fig:Biaxial_schematics}c, we see a gradual drop of the
maximum global $g^{2}$. Although the system is effectively ``jammed"
for part of each cycle, after enough cycles, the stress relaxes to a
measurably lower value.  We then track the evolution of these jammed
states by considering $g^2$ for the most compressed states. The
corresponding time series of $g^{2}/g_0^2$, for several packing
fractions of ellipses and disks (the most compressed $\phi$ is used as
a label here), are shown in Fig.~\ref{fig:gsqr_time_series_2}a. Here,
we emphasize the relaxation which occurs for states with a
most-compressed packing fractions that are above the isotropic jamming
point.  However, there is a largest $\phi_m$ such that we do not see
significant particle or stress relaxation over $~\sim 1000$ cycles.
For instance, the time series of disks corresponding to $\phi = 0.868$
in Fig.~\ref{fig:gsqr_time_series_2}a pertain to a packing fraction
which is lower than $\phi_m^{disks}$, even though the fully compressed
density for this data is above the isotropic jamming density of $\phi
\simeq 0.84$.  At a higher packing fraction (i.e. $\phi = 0.883$), the
global $g^{2}$ is persistent for many cycles, in the sense that there
is no overall change in the peak $g^2$ with cycle number.  We conclude
that there is a range of $\phi$ for which the very small amount of
spatial freedom, coupled with the compressive/dilational driving is
sufficient to allow structural rearrangement.

In order to quantify the long term stress relaxation as in
Fig.~\ref{fig:gsqr_time_series_2}a, we fit the time series to the
functional form $\frac{g^2}{g^2_0}= A (\frac{t_0}{t}+1)^{a}$, which
captures both the initial relatively fast drop and the long-time
saturation to a final value, $g^2_{\infty}/g^2_0$. Since, we have one
more constraint on parameters at $t=1$, i.e. $A(t_0+1)^a=1$, the
functional form reduces to
$\frac{g^2}{g^2_0}=(\frac{t+t_0}{t(t_0+1)})^{a}$. The fitted curves
are shown by solid lines in Fig.~\ref{fig:gsqr_time_series_2}a. The
large time value is $g^2_{\infty}/g^2_0 = (1 + t_0)^{-a}$, and it is
straight forward to compute the time $t_{1/2}$ for $\frac{g^2}{g^2_0}$
to fall from 1.0 to $(1+ g^2_{\infty}/g^2_0)/2$.

From the parameters $a$
and $t_0$, obtained from the least squares fits, we find
$g^2_{\infty}/g^2_0$ and $t_{1/2}$ as a function of density for both
disks and ellipses, which we then show in
Fig.~\ref{fig:gsqr_time_series_2}b,c. Although there is a fair bit of
scatter, the time scale $t_{1/2}$ grows strongly and
$g^2_{\infty}/g^2_0$ jumps quickly to 1.0, above a characteristic
maximal packing fraction, $\phi_m$, where the system becomes
effectively frozen; at least on the time of these experiments, there
is little evolution of the 
global stress for either ellipses or disks
for cyclic compression carried out above this density.  
From $g^2_{\infty}/g^2_0$ values, 
We estimate $\phi_m \simeq 0.88$ for disks, and $\phi_m \simeq
0.93$ for ellipses. For densities above $\phi_m$, the system maintains
a global $g^2$ and a memory of its previous state for seemingly
arbitrarily large numbers of compression cycles.  To the best of our
knowledge, this work is the first observation of such an effect.

\begin{figure}
\centering
{\includegraphics[scale=0.43]{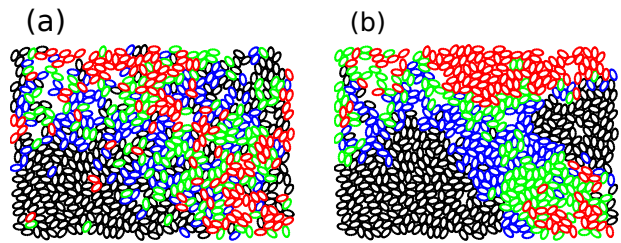}}
\caption{ Mobility of the particles in an arbitrary time. The packing
  fraction of the system is $\phi=0.932$. The time delay, $\tau$, equals
  a) 10 cycles, and b) 1000 cycles. Different colors represent
  particles which have displaced more than $A\%$ and less than $B\%$
  of all particles. Where we have red: $A=80$ \& $B=100$, green:
  $A=60$ \& $B=80$, blue: $A=40$ \& $A=60$, and black: $A=0$ \&
  $A=40$.}
\label{fig:spatial_heterogeneity}
\end{figure}

{\it Particle motion vs. stress evolution:} Given that the present systems
can compact over long time scales, an interesting question is whether
the dynamical process of gradual compaction is associated with
identifiable, and possibly heterogeneous, structural changes (i.e. in
the particle positions).  In order to probe the effect of particle
motion, we first consider the mobility of particles, 
defined as the displacement of the
particle for a given time delay, $\tau$, relative to the mean
displacement of all particles (Here, time represents the number of compression cycles).
Fig.~\ref{fig:spatial_heterogeneity} shows data for two different time
delays, $\tau=10$, and $\tau=1000$ in a compression experiment on ellipses.  Particles with similar mobility are represented
by similar colors in Fig.~\ref{fig:spatial_heterogeneity}
Although the spatial distribution of
highly mobile particles change substantially over time, the regions of
comparable mobility form very large clusters. This indicates heterogeneous dynamics both in time and
space. The dense structure of these clusters suggests that there are at best small local rearrangements of the particles.

\begin{figure}
\centering
{\includegraphics[scale=0.37]{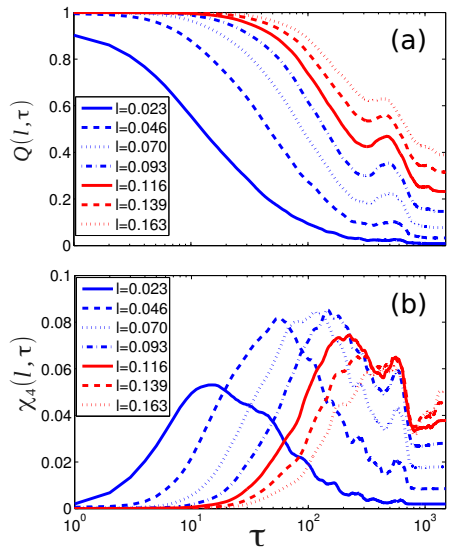}}
\caption{$Q_s(\tau)$ and $\chi_4 (\tau)$ for a system of isotropically
  compressed ellipses with (highest) packing fraction of
  $\phi=0.916$. The unit of length scale, $l$, is the semi-minor axis
  of ellipses.}
\label{fig:X4_Q}
\end{figure}

As an alternative approach of
quantifying this heterogeneous dynamics, we have studied the 4-point
susceptibility, $\chi_{4}(\tau)$, which indicates the extent of
temporal correlation of dynamics at any pair of spatial points\cite{keys07}. $\chi_{4}(\tau)$ is defined as:
$\chi_{4}(\tau)=N[<Q_{s}(\tau)^{2}>-< Q_{s}(\tau)>^{2}]$.  We choose
$Q_{s}(t)=\frac{1}{N} \sum_{i=1}^{N} w(|r_{i}(t)-r_{i}(0)|)$, with
$$
w= \left\{ \begin{array}{rl}
 1 &\mbox{ if  $|r_{i}(t)-r_{i}(0)|<l$}, \\
 0 &\mbox{  otherwise}
       \end{array} \right.
$$
\noindent
$N$ is the number of particles, $r_i(t)$ indicates the particle
positions at time $t$, and $l$ is a length scale which is an
adjustable parameter. The averages are taken over all the particles
and over all starting times. $Q_s(\tau)$, which is referred to as the
\emph{self overlap order parameter}, is a measure of
particle mobility, and is quantified by a length scale $l$\cite{keys07}.  We plot
$Q_s(\tau)$ and $\chi_4 (\tau)$ vs. $l$, in Fig.~\ref{fig:X4_Q}. As
seen, $Q_s$ varies from $1$ to $0$ as the time delay $\tau$
increases. On the other hand, $\chi_4$ has a maximal point for each
$l$, which basically characterizes a time delay, $\tau^*$, by which
the particles on average move more than the length scale $l$.

\begin{figure}
{\includegraphics[scale=.5]{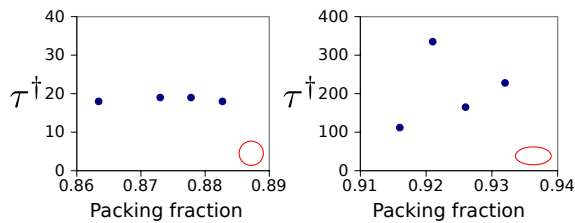}}
\caption{$\tau^{\dag}$ vs. packing fraction, $\phi$,
  for systems of ellipses and disks. As seen, the average value of
  $\tau^{\dag}$ is about an order of magnitude larger for ellipses
  compared to disks. }
\label{fig:tau_dagger}
\end{figure}

As seen in Fig.~\ref{fig:X4_Q}b, $\chi_4$ is maximized for
$l^{\dag}\simeq 0.093$, where $l^{\dag}$ is the characteristic length
scale. We now take the characteristic $l^{\dag}$ for each packing
fraction, and find the corresponding maximal $\tau^{\dag}$. The data
are demonstrated in Fig.~\ref{fig:tau_dagger}. We note that similar
structure for $\chi_4$ has been obtained under much more
energetic driving conditions by Dauchot and coworkers\cite{Dauchot05}.
There are several remarkable features in these data.  First, the
typical length scales for $l$ are only a fraction of a particle
diameter; the particles are largely confined.  Second, the
characteristic times $\tau^{\dag}$ are relatively insensitive to
$\phi$, but are an order of magnitude greater for ellipses than for disks.  

\begin{figure}
{\includegraphics[scale=.033]{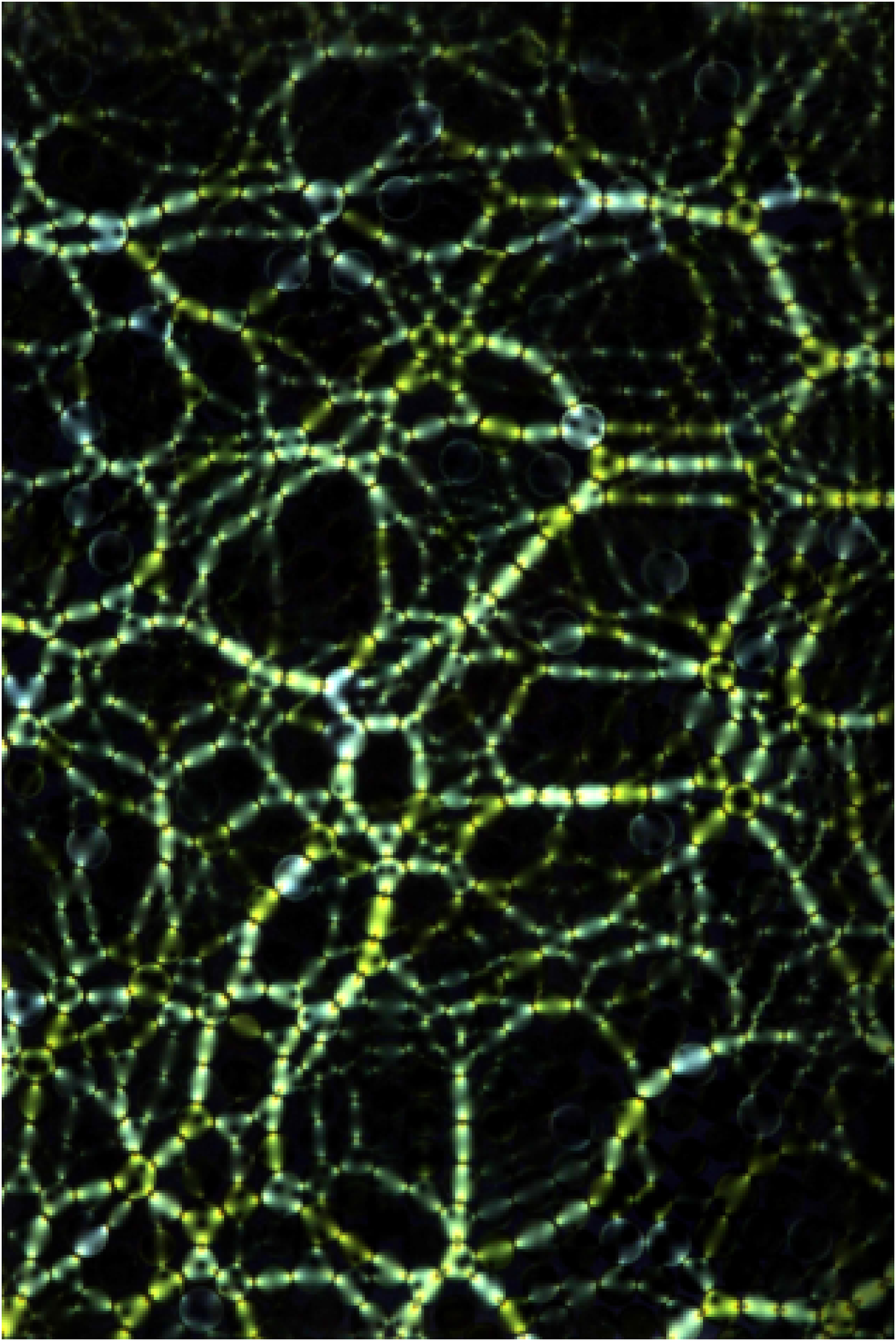}}
{\includegraphics[scale=.033]{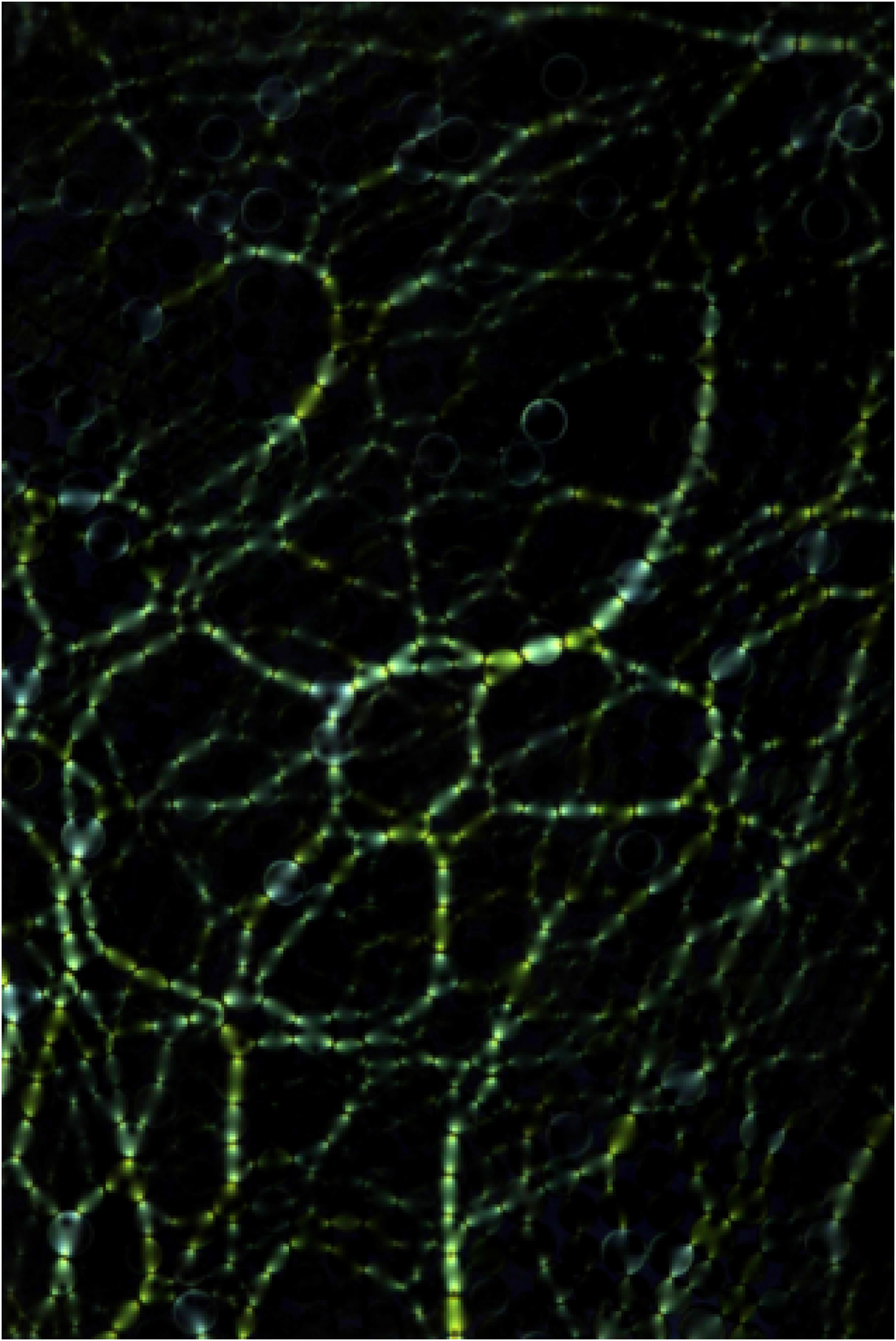}}
{\includegraphics[scale=.033]{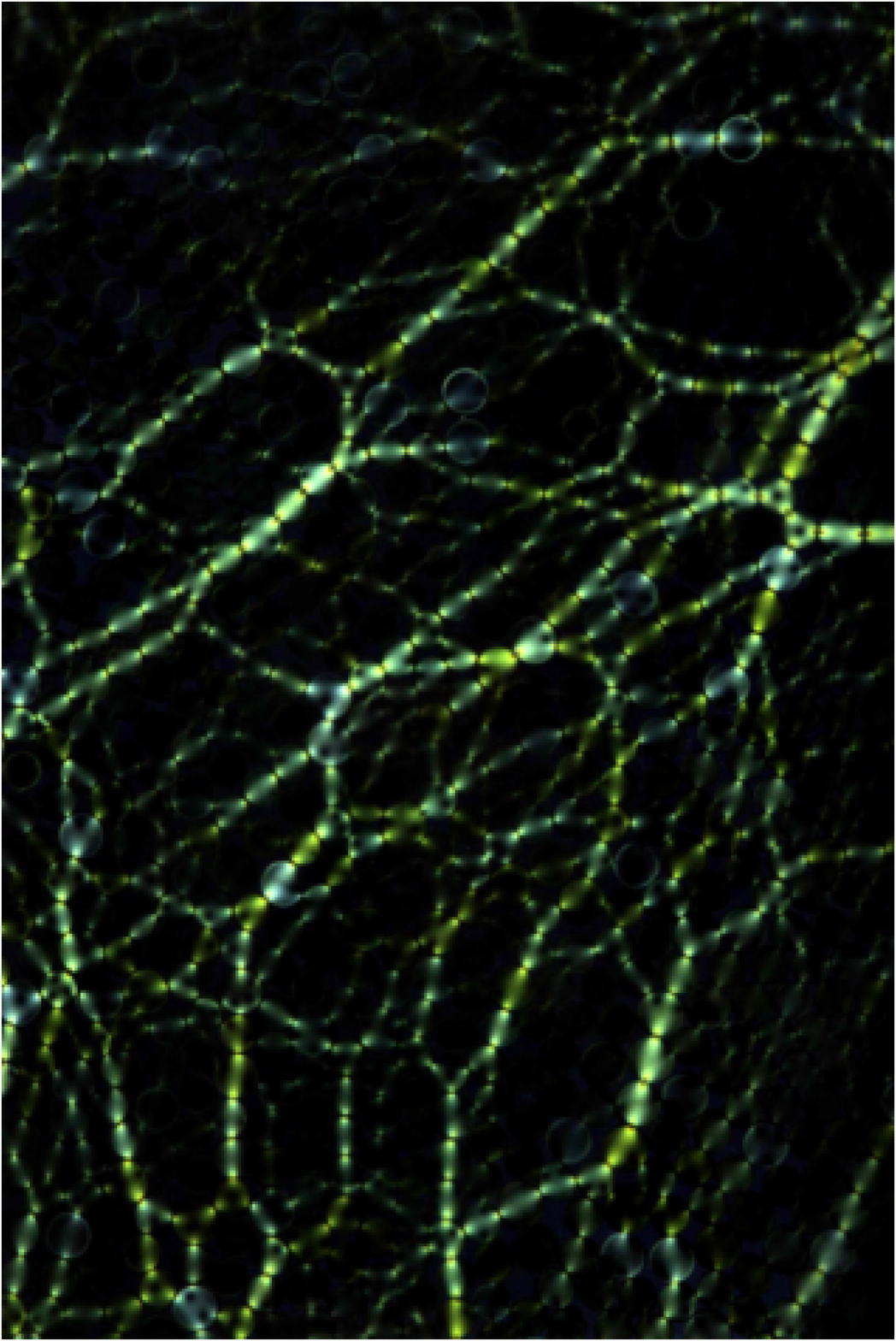}}
\caption{Photoelastic images showing how the force network evolves
  over 999 cycles of cyclic compression, applied to a system of disks
  at $\phi = 0.863$. The cycle number of images from left to right are: 1, 500, and 999.}
\label{fig:force-networks}
\end{figure}

These results beg the question, where should one look to observe the
stress relaxation demonstrated in Fig.~\ref{fig:gsqr_time_series_2}a.
In fact, Fig.~\ref{fig:force-networks} shows that although the
particles may move very little, even a tiny bit of freedom allows the
force network to evolve and relax substantially.  In Fig.~\ref{fig:force-networks}, we
show photoelastic images of disks for the the maximum compressions of
cycle 1, 500 and 999.  In fact, there are substantial changes in these
networks.

{\it Conclusions:} We have observed transient stress states that occur
for both systems of disks and ellipses that are cyclically compressed
by a modest amount above the isotropic jamming point (point $J$). The
global stress relaxes to a stationary value in the course of many
compression cycles. 
The time corresponding to stress relaxation
grow substantially above a
characteristic packing fraction, $\phi_m$ (ellipses: $\phi_m=0.93$;
disks: $\phi_m=0.88$). We have sought to identify the origin of the
stress relaxation.  To do so, we first characterized the structural
changes by computing relative mobilities and, $\chi_4$.  In
particular, the characteristic time scale, $\tau^{\dag}$, varied
rather little with $\phi$ whereas the stress relaxation showed a
significant change by increasing $\phi$. 
However, the analysis of $\chi_4$ did identify significantly longer
times for $\tau^{\dag}$ in the case of ellipses, which we attribute to
the fact that ellipses are substantially confined by their inability
to rotate.  In fact, the most clearly identifiable relaxation occurs
in the force network, even though there is minimal particle motion.
An interesting issue for future work concerns the extent to which
inter-particle friction plays a role in the relaxation process.  For
disks, friction may be more important in stabilizing packings than for
ellipses, where rotational constraints do occur even without friction.

{\it Acknowledgement:} This work is supported by NSF grant
DMR-1206351, and ARO grant W911-NF-1-11-0110.

\end{document}